\documentclass[10pt,fleqn,english]{article}
\usepackage[]{geometry}
\usepackage{braket}
\usepackage[utf8]{inputenc} 
\usepackage{amsmath}
\DeclareMathOperator{\Tr}{Tr}

\usepackage{siunitx}
\usepackage{amsfonts}
\usepackage{amssymb}
\usepackage[super]{nth}
\usepackage{csquotes} 
\usepackage{hyperref}
\usepackage{mathtools}
\usepackage{lmodern}
\usepackage{amsfonts}
\usepackage{microtype}
\usepackage{xcolor}
\usepackage{booktabs}
\usepackage{graphicx}
\usepackage{moreverb}
\usepackage{multicol}
\usepackage{float,gensymb}

\usepackage[font={small,md},labelfont=bf]{caption}

\newenvironment{Figure}
  {\par\medskip\noindent\minipage{\linewidth}}
  {\endminipage\par\medskip}

\usepackage{scrpage2}
\usepackage[T1]{fontenc}
\usepackage{gensymb}
\usepackage{tabu}

\usepackage{soul}

\ofoot{\pagemark}
\graphicspath{{ressources/}}

\usepackage{tikz}
\usepackage[backend=bibtex,style=nature,maxbibnames=5]{biblatex}
\bibliography{bibliography.bib} 

\usepackage{notoccite}

\begin{document}

\newcommand{\newtitle}[1]{
\begin{center}
{\textbf {\fontsize{14pt}{16.8pt}\selectfont #1}}
\end{center}
}

\newcommand{\authors}[1]{
\begin{center}
\textbf{\textit{\fontsize{12pt}{14.4pt}\selectfont #1}}
\end{center}
}

\newcommand{\newaffiliations}[1]{
\begin{center}
\textit{\fontsize{8pt}{8pt}\selectfont #1}
\end{center}
}


\thispagestyle{empty}

\newtitle{Experimental Quantum Homomorphic Encryption} 

\authors{Jonas Zeuner$^1$*, Ioannis Pitsios$^{2}$, Si-Hui Tan$^{4,5}$, Aditya N. Sharma$^{1}$, Joseph F. Fitzsimons$^{4,5,6}$, Roberto Osellame$^{2,3}$ and Philip Walther$^{1,6}$ }
\newaffiliations{
$^1$Vienna Center for Quantum Science and Technology, Faculty of Physics, University of Vienna, Boltzmanngasse 5, Vienna A-1090, Austria.\\
$^2$Istituto di Fotonica e Nanotecnologie - Consiglio Nazionale delle Ricerche (IFN-CNR), p.za Leonardo da Vinci 32 20133, Milano, Italy\\
$^3$Department of Physics - Politecnico di Milano, p.za Leonardo da Vinci, 32, 20133 Milano, Italy\\
$^4$Singapore University of Technology and Design, 8 Somapah Rd, Singapur 487372\\
$^5$Centre for Quantum Technologies, National University of Singapore, 3 Science Drive 2, Singapore 117543\\
$^6$Erwin Schrödinger International Institute for Mathematics and Physics, Boltzmanngasse 9A, 1090 Wien, Austria
}

\begin{changemargin}{1cm}{1cm} 
\noindent
\begin{small}
Quantum computers promise not only to outperform classical machines for certain important tasks \cite{feynman1982}, but also to preserve privacy of computation. For example, the blind quantum computing protocol \cite{bqc_theo,barz2012demonstration} enables secure delegated quantum computation, where a client can protect the privacy of their data and algorithms from a quantum server assigned to run the computation. However, this security comes at the expense of interaction: the client and server must communicate after each step of the computation. Homomorphic encryption, on the other hand, avoids this limitation. 
In this scenario, the server specifies the computation to be performed, and the client provides only the input data, thus enabling secure non-interactive computation \cite{gentry,vanDijk2010,pointcheval2010public,broadbent2015quantum,dulek2016quantum,mahadev2017classical,lai2017statistically,ouyang2015quantum,newman2017limitations}. 
 Here we demonstrate a homomorphic-encrypted quantum random walk using single-photon states and non-birefringent integrated optics. The client encrypts their input state in the photons' polarization degree of freedom, while the server performs the computation using the path degree of freedom \cite{rohde2012quantum}. Our random walk computation can be generalized, suggesting a promising route toward more general homomorphic encryption protocols \cite{homo2016quantum}.
 \end{small}
\end{changemargin}

\vspace{0.5cm}
\noindent
Secure delegated computing has been a longstanding research goal for both the classical and quantum computation communities. The aim is to provide a client (Alice) access to remote computational resources (Bob), while protecting the privacy of the data or the algorithm.
In his seminal 2009 paper, Gentry described the first computationally secure, fully homomorphic encryption scheme for classical computing \cite{gentry}. Here, \enquote{computational security} means that the privacy guarantees of the protocol are based on assumptions about an adversary's computational capabilities; \enquote{fully} means that  \textit{any} computation is possible.
Blind quantum computation was as well introduced in 2009 \cite{bqc_theo,barz2012demonstration}, enabling a client to protect both their data and algorithm while running arbitrary computations on a remote quantum computer. 
While blind quantum computation has the  important advantage of being  information-theoretically secure --- i.e., it does not rely on assumptions about the adversary's technological capabilities --- its efficiency is limited by the need for interaction: 
Alice and Bob must exchange classical information after each step of the computation.
Quantum homomorphic encryption removes the requirement of interactive computation but necessarily sacrifices either security or computational power to achieve this, in accordance with a no-go theorem \cite{nogotheorem}: fully homomorphic encryption is impossible if perfect privacy and non-exponential resource overhead are required. Therefore, in the protocol implemented here, the requirements for (1) universal computation and (2) perfect privacy are relaxed, based on the following two observations \cite{rohde2012quantum}. (1) Certain classes of computations, such as quantum random walks, while only subsets of universal quantum computation, are nevertheless of great interest \cite{aaronson2011computational,tillmann2013experimental,spagnolo2014experimental,broome2013photonic,carolan2015universal}. (2) In any practical encryption application \textit{perfect} privacy is not required, as long as the maximum amount of information potentially available to an attacker is sufficiently small.

\begin{Figure}
 \centering
 \includegraphics[width=0.6\linewidth]{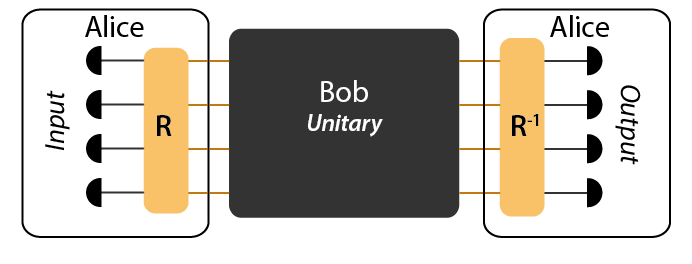}
 \captionof{figure}{\textbf{Homomorphic encryption scheme.} Alice prepares her input state by encoding the desired photon-number state in the $\{H,V\}$ basis and then encrypting it by applying a randomly chosen polarization transformation on all photons ($R$). Bob performs the  quantum computation on the encrypted state and returns the photons to Alice. Alice undoes the previous transformation ($R^{-1}$)  and measures the photons in the $\{H,V\}$ basis, obtaining the outcome of the quantum computation.}
  \label{fig:homo_sketch}
\end{Figure}
\noindent

In this experiment, we use single-photon qubit input states and an integrated-optics server to demonstrate a quantum random walk using homomorphic-encrypted data, as proposed in \cite{rohde2012quantum}. Quantum random walks are typically implemented with either $0$ or $1$ photon in each input mode, distributing $n$ photons over $m$ spatial modes.  The protocol we implement here instead uses the photons' polarization to encode Alice's input state for the quantum random walk, taking advantage of the fact that orthogonally polarized photons do not interfere.

\begin{figure*}[h]
\center
\includegraphics[width=0.8\textwidth]{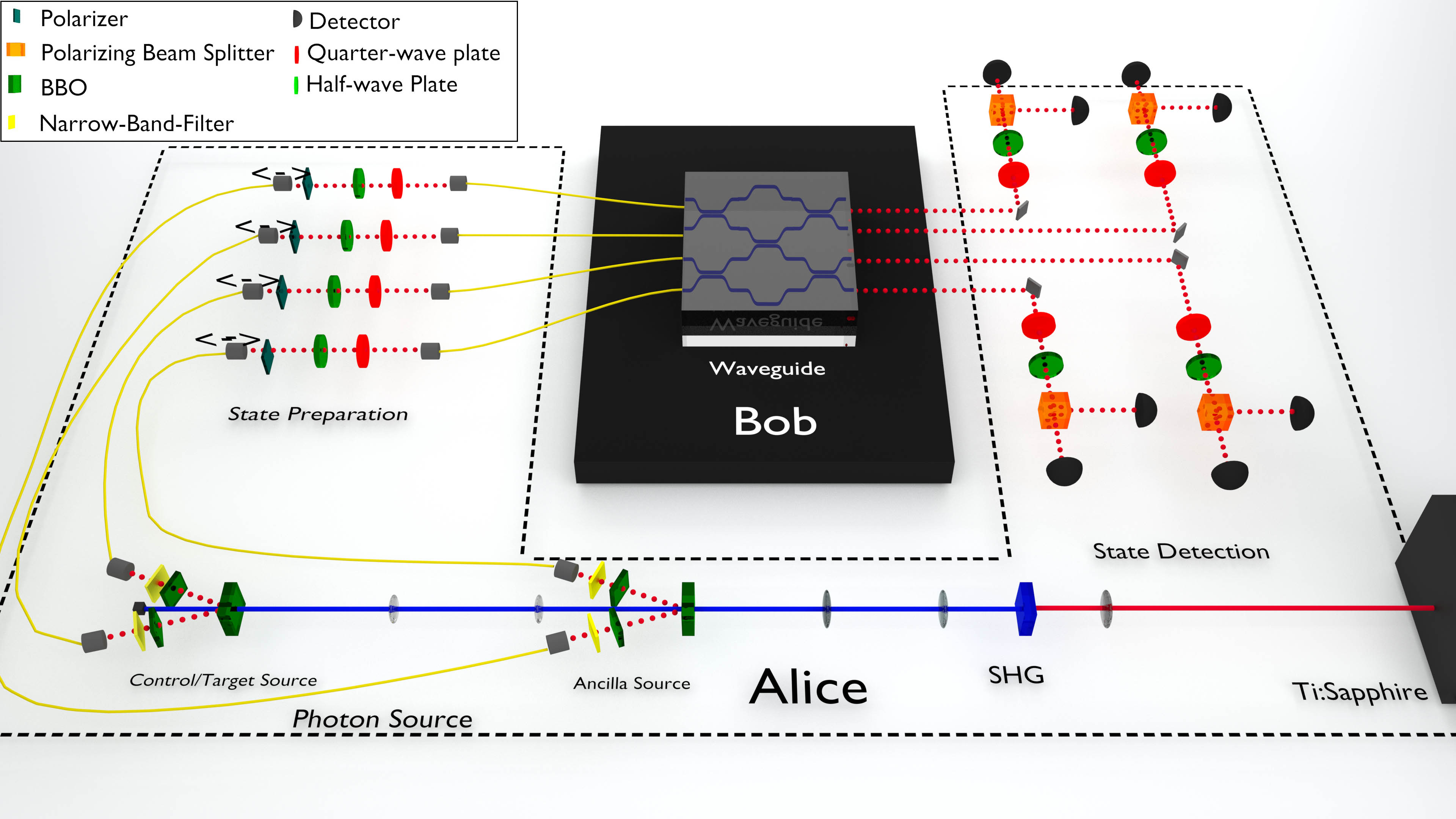}
\caption{\textbf{Experimental setup.} A Ti:Sapphire laser is used to pump two non-linear $\beta$-barium borate crystals, each probabilistically producing exactly one pair via type-II spontaneous parametric down conversion. These photons are spectrally filtered and sent through polarizers to prepare a pure, separable four-photon state. The four photons are coupled to single-mode fibers and synchronized in the delay stage, using adjustable free-space delays (indicated by the double arrows). Using half-wave plates (HWPs) and quarter-wave plates (QWPs) Alice can prepare arbitrary polarization states before sending the photons to Bob, who will perform the random walk.
After exiting Bob's chip, the four output modes are collimated by a lens and sent back to Alice. She uses the detection stage (HWP, QWP,  polarizing beam splitter (PBS), single-photon detector for each photon) to projectively measure the photons and recover the outcome of the random walk.}
\label{fig:setup}
\end{figure*}

\noindent
Thus, to implement an $m$-mode quantum random walk of $n$ \enquote{walker} photons, rather than inputting one photon into each of $n$ modes and leaving the remaining $m-n$ empty, we also input $m-n$ \enquote{dummy} photons in the otherwise empty modes, with polarizations orthogonal to the $n$ photons representing the walkers.
For example, an input state $\ket{\Psi_{in}}=\ket{1,0,0,0}$ for a traditional quantum random walk (written in the occupation-number basis) would be encoded in this scheme as $\ket{\Psi_{in,encoded}}=\ket{H,V,V,V}$, where $\ket{H}$ $(\ket{V})$ represents horizontal (vertical) polarization. Measuring the output photons in the $\{H,V\}$ basis then yields the same result as the traditional occupation-number quantum random walk. 
The purpose of this approach is to enable polarization encryption of Alice's input state:  without knowing the basis in which Alice's input is encoded, Bob can guess Alice's input state with only limited probability of success.
To encrypt the input state $\Psi_{in}$, Alice randomly chooses a key, a polarization state $\ket{X}$ taken from a set of $d$ uniformly distributed points on the Poincar\'e sphere, where $d$ is the number of polarization basis choices available to her. To encrypt her data, Alice rotates the polarizations of her qubits from $\ket{H}$ and $\ket{V}$ to $\ket{X}$ and $\ket{X^\perp}.$ 
Alice sends this encrypted state to Bob, who performs the quantum random walk. Bob returns the output photons to Alice, and she measures them in the $\{X,X^\perp\}$ basis, obtaining the result of the random walk.

\begin{figure*}[h]
\center
\includegraphics[width=0.9\textwidth]{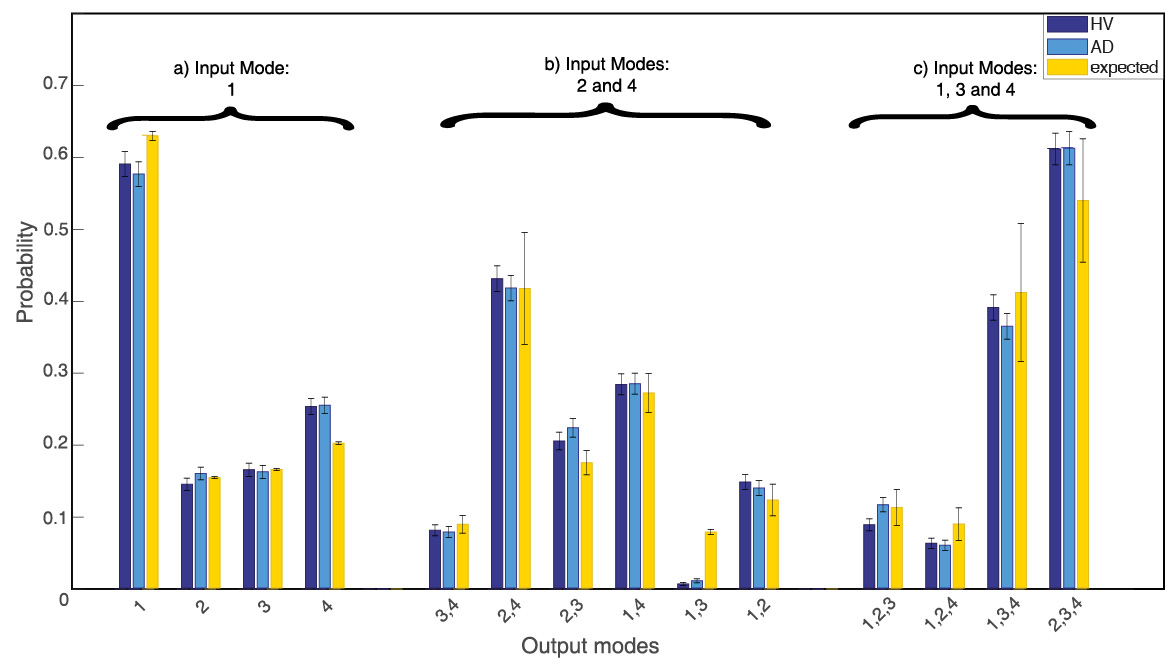}
\caption{\textbf{Results of the encrypted random walk.} We use two different devices to execute multiple encrypted random walk computations  (see  Figure 5 for the equivalent results on the second device). (a) One walker, (b) two walker, and (c) three walker are launched into different input modes alongside a corresponding number of dummy photons. Here the output probabilities for three different input cases for two mutually unbiased polarization bases are shown, alongside the values expected based on the reconstructed unitary (see Appendix). The fidelities (Bhattacharyya distance \cite{bhattacharyya1943measure}) between the expected and both measured probabilities are (a) $0.99\pm0.02$, (b) $0.99\pm0.02$ and (c) $0.99\pm0.03$ and demonstrate the polarization independence of the computation. More data and discussion of error analysis is provided in the  Appendix.}
\label{fig:results}
\end{figure*}

\noindent
If Bob tried to decipher Alice's encrypted state, the amount of information he could extract is bounded by the Holevo quantity \cite{holevo1973bounds}. One straightforward attack Bob could employ is to randomly choose a basis in which to measure all $m$ photons: in fact, this attack is close to optimal, almost saturating the Holevo bound.  In the limit of large $d$ and $m$, the success probability of this attack is $p_B=1/\sqrt{\pi m}$ \cite{rohde2012quantum}.
The protocol also ensures the privacy of Bob's algorithm.  Since Alice only knows the input and output states of the computation, the amount of information that she can extract about Bob's algorithm is proportional to that of a \enquote{black-box} function:  the more queries she is allowed to send, the more accurately she can guess the function. 
It is important to note that both Alice and Bob have an interest in performing a certain computation on a certain input state exactly once, since both of them increasingly compromise the privacy of their respective secrets with increasing number of repetitions of the computation. The no-go theorem \cite{nogotheorem} asserts that this limitation is unavoidable.

In our experimental demonstration, Alice produces four photons using two spontaneous parametric down-conversion (SPDC) sources (see Appendix) and prepares them in a randomly chosen polarization state using a polarizer, half-wave plate (HWP), and quarter-wave plate (QWP) for each photon. Alice can create input states of any polarization with a fidelity of $(99.5\pm0.1)\%$, the main source of error being imperfect polarization compensation of the single-mode fibers leading to the chip. 
After preparing the encrypted input state, Alice sends the photons to Bob, who performs the random walk.

In order for the scheme to work, Bob's chip must implement the same unitary for the photons' path degree of freedom regardless of the input polarizations used --- otherwise, the outcome would depend on Alice's choice of key. Although laser-written waveguides support propagation of all polarizations, they typically have slightly different refractive indices for $H$ and $V$ polarizations ($\Delta n \approx 10^{-5}$), making it a challenge to implement nontrivial polarization-independent path unitaries. To achieve this, we used an annealing procedure to fabricate waveguides with birefringences $\Delta n < 10^{-6}$ (see Appendix).

After the random walk, Bob returns the photons to Alice, who projects them in her previously chosen polarization basis using QWPs, HWPs, polarizing beam splitters (PBSs), and single-photon detectors.
To demonstrate the fidelity of the homomorphic-encrypted random walk we chose a canonical set of two mutually unbiased polarization bases and performed random walks with one, two, and three walkers using two different unitaries, each with $m=4$ inputs and outputs.
We used $\{H,V\}$ (parallel and orthogonal, respectively, to the chip surface) and $\{D,A\}$  ($\ket{D}=\frac{1}{\sqrt{2}}(\ket{H}+\ket{V})$ and  $\ket{A}=\frac{1}{\sqrt{2}}(\ket{H}-\ket{V})$).
We characterized the unitary and compared the output probability distributions with theoretical predictions, finding the mean overlap (Bhattacharyya distance \cite{bhattacharyya1943measure}) between the predictions and results from all random walks to be  $(0.995\pm0.014 )\%$ for the first device (Fig. \ref{fig:results}) and $(0.986\pm0.012)\%$ for the second (see Appendix). 

The security guarantees for Alice’s plain-text input state can be quantified in various ways. The trace distance between the different input states that she can produce with four photons is 0.81 for Hamming distances 1 and 3 and 0.85 for Hamming distance 2 \cite{hamming1950error}. As a result, Bob cannot perfectly distinguish any pair of possible plain-texts. Furthermore, the mutual information between her plain-text string and Bob is bounded by the Holevo quantity to be no more than $1.96$ bits (see Appendix). To experimentally verify the security of Alice's input, we implemented the attack described above: Bob measures all of Alice’s four photons in a randomly chosen basis (here we choose $\ket{H}$ for simplicity). Alice encrypts her plain text input-state  (here we use $\ket{1,1,1,1}\equiv\ket{H,H,H,H}$) by choosing between  $d=2,3,4,6,12$ different linear polarization bases (keys). The probability of Bob guessing Alice's plain-text input-state can then be determined from the fraction of four-fold coincidence detections Bob measures with polarization $\ket{H,H,H,H}$ (see Fig.~\ref{fig:random_attack}). For $m=4$, this probability asymptotically approaches $p=0.27$ for large  $d$. Note that with current technology it is already straightforward, in principle, to achieve arbitrarily large $d$, although this leads to only a minor improvement: Alice could \enquote{auto-calibrate} her encrypting and decrypting operations by using a single QWP, HWP set to both prepare and measure all of her photons' polarizations. Improving Alice's security requires increasing $m$ (for example, $p_B=0.01$ for $m=3500$). Additionally, Alice can reduce the Holevo information by a factor of 2 by selecting keys from the whole Poincar\'e sphere, rather than constraining herself to linear polarizations.

\begin{Figure}
 \centering
 \includegraphics[width=0.6\linewidth]{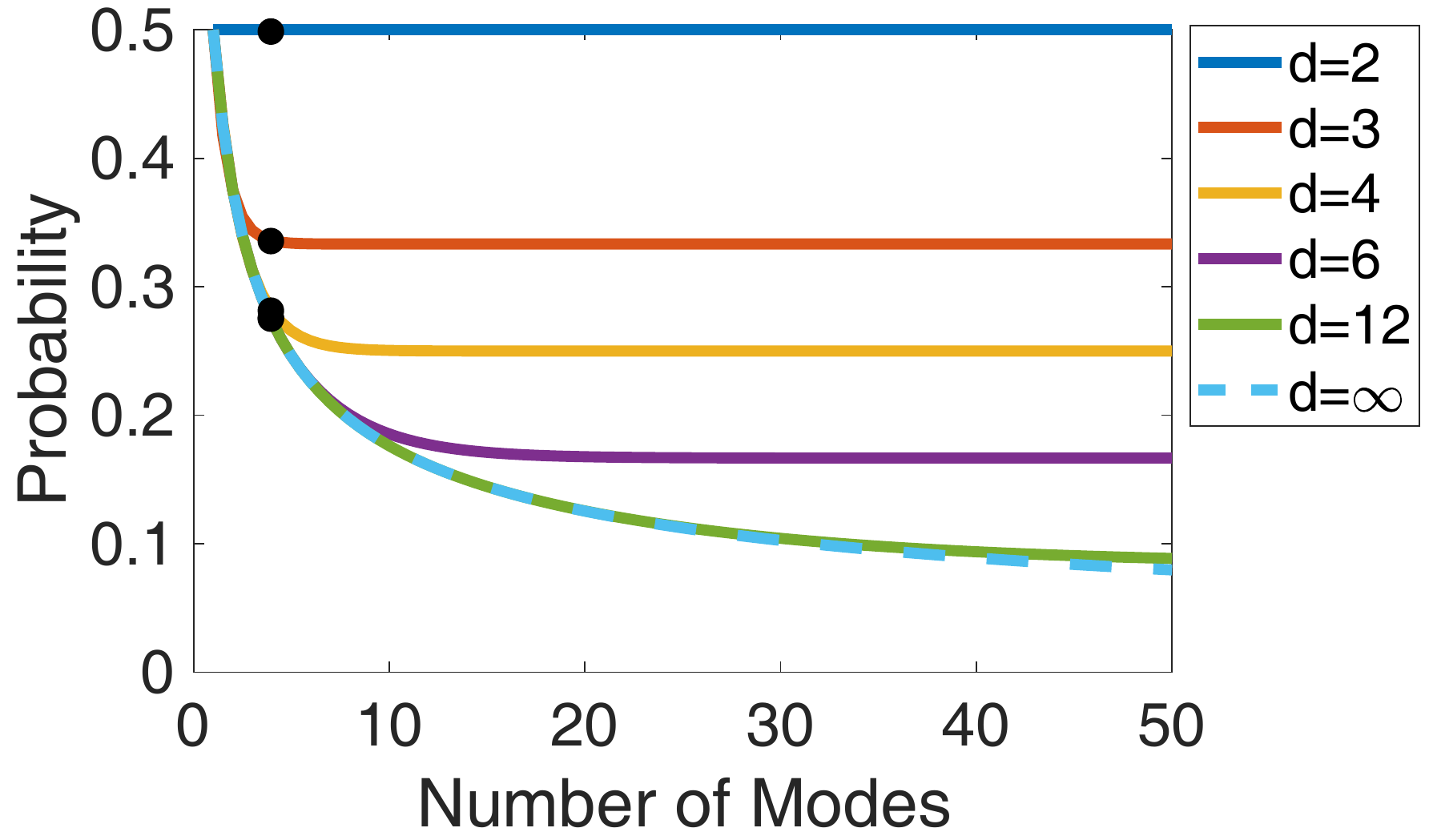}
 \captionof{figure}{\textbf{Privacy of Alice’s plain text input state.} In our homomorphic encryption scheme, a random attack is optimal when Alice encrypts with all polarizations, and nearly optimal when she encrypts with only linear polarizations: Bob measures all photons in a randomly chosen basis. Alice’s security depends on the number of polarization bases (keys) $d$ she can choose from, and the input modes $m$. In our case $m=4$ and we measure the probability of Bob guessing the correct state for $d=2,3,4,6,12$ (black dots, error bars lie within the points). The blue dashed line shows the asymptotic behaviour for an infinite key number. All lines are theoretical upper bounds (see Appendix).}
 \label{fig:random_attack}
\end{Figure}

\noindent
We have demonstrated homomorphic-encrypted quantum random walks of up to three walkers in four modes. Our photonic system's specially engineered features allowed us to encrypt Alice’s plain-text input state in polarization, while performing computations using the path degree of freedom. The security of Alice’s plain-text input is necessarily limited by the number of modes used, i.e. by the number of available photons --- however, the continuing advances in photon-source technology will enable similar demonstrations using more modes in the future. Further improvements can be made by encrypting in a different photonic degree of freedom with more than two levels. For example, orbital angular momentum enables, in principle, arbitrarily high-dimensional encoding, and transmission of such states in optical fiber has already been demonstrated \cite{Bozinovic1545}. Using an $a$-level degree of freedom for encoding, instead of polarization, the amount of hidden information can be improved from $\log_2(m)$ scaling to $m \log_2(a/m) +m (\log(2))^{-1}$ \cite{homo2016quantum}. As we have shown here, although \textit{perfect} security for \textit{universal} computation (without exponential resource overhead) is forbidden \cite{nogotheorem}, relaxing these conditions can enable interesting applications. Determining the ideal mix of security, performance, and generality of the computation remains an active topic of research.

\subsection*{Data availability}
The authors declare that the main data supporting the finding of this study are available within the article and its Appendix.
Additional data can be provided upon request.

\subsection*{Acknowledgements}
P.W. acknowledges support from the research platform TURIS, the Austrian Science Fund (FWF) through the Doctoral Programme CoQuS (no. W1210-4) and NaMuG (P30067-N36), the United States Air Force Office of Scientific Research via QAT4SECOMP (FA2386-17-1-4011) and Red Bull GmbH.

R.O. acknowledges support from the  ERC-Advanced Grant CAPABLE (Composite integrated photonic platform by femtosecond laser micromachining; no. 742745).

P.W. and R.O. acknowledge support from the European Commission via Photonic Integrated Compound Quantum Encoding (PICQUE) (no. 608062) and Quantum Simulation on a Photonic Chip (QUCHIP) (no.
641039) projects.

S.T. and J.F. acknowledge support from Singapore's National Research Foundation under NRF award NRF-NRFF2013-01 and from the United States Air Force Office of Scientific Research under grant no. FA2386-15-1-4082.

\subsection*{Competing interests}
J.F. has financial holdings in Horizon Quantum Computing Pte Ltd.

\subsection*{Author contributions}
J.Z, A.S. built the setup and carried out data collection, J.Z., A.S. performed data analysis, I.P. designed and fabricated the chip, S.T. and A.S. contributed to the theoretical calculations, J.F., R.O. and P.W. supervised the project. All authors contributed to writing the paper.

\subsection*{References}
\printbibliography[heading=none]

\newpage
\section*{Appendix}

\subsection*{Experimental Setup.} Our experimental setup is shown in Figure \ref{fig:setup}. We generate all four photons using degenerate, noncollinear type-II spontaneous parametric down-conversion (SPDC). Two separate $\SI{2}{mm}$-thick $\beta$-barium borate (BBO) crystals are pumped by a Ti:Sapphire laser (Coherent Chameleon Ultra II, $\SI{789}{nm}$, $\SI{150}{fs}$ duration, $\SI{80}{MHz}$ repetition rate, $\SI{3.6}{W}$ average power) which has been frequency doubled to $\SI{394.5}{nm}$ using second harmonic generation in a $\SI{5}{mm}$-thick lithium triborate (LBO) crystal. The photons emitted by the crystals pass through  $\SI{1}{mm}$-thick BBO crystals of the same cut angle as the SPDC crystals to compensate for spatial and temporal walk-off before being spectrally filtered by $\SI{3}{nm}$-bandwidth spectral filters centered at $\SI{789}{nm}$, and spatially filtered by single-mode optical fibers (SMFs) of type Nufern 780-HP.
All photons pass through polarizers to create pure polarization states and then through a half-wave-plate (HWP) and quarter-wave-plate (QWP) to enable the creation of arbitrary polarizations states. The QWP and HWP were rotated using highly precise motorized rotation mounts with a precision of $0.02^{\degree}$.
Adjustable free-space delay lines are used to synchronize the photons such that they all arrive at the chip within their coherence time of approximately \SI{300}{fs}. The photons are coupled to the chip using a \SI{127}{\micro\meter}-pitch v-groove array of Nufern 780-HP fibers. The \SI[product-units=brackets]{5x 5}{\micro m}  fiber mode-field has a high overlap with the mode-field of the waveguides, which are of equivalent size. On the output facet of the chip the photons are collimated using a lens and sent to the detection stage. Using a QWP, HWP and a polarizing-beam-splitter (PBS) and avalanche photodiodes (APDs) the photons can be detected in any desired polarization basis.
The overall transmission (from fiber in-coupling to APDs) was measured to be ($50\pm 5$)\%. 

\noindent
\subsection*{Waveguide Details.}
The four-mode optical circuit for our quantum random walk was fabricated by direct laser writing in Corning Eagle-XG borosilicate glass. The laser source we employed was a Yb:KYW cavity-dumped oscillator at \SI{1030}{nm} wavelength, emitting pulses of \SI{300}{fs} duration, and at \SI{1}{MHz} repetition rate. The laser beam was focused into the bulk of the glass substrate using a 50x, \SI{0.6}{NA} microscope objective, and the inscription of the optical waveguides was performed by translating the glass (with respect to the objective’s focus), with a computer-controlled three-axis Aerotech FiberGlide 3D series stage, at a tangential velocity of \SI{40} {mm/s}. The waveguides were inscribed at a depth of \SI{170}{\micro m}, with \SI{270}{mW} of laser power, using a multiple irradiation approach (5-times per waveguide), and then they were annealed.  The thermal processing makes the optical circuits polarization insensitive \cite{corrielliAnnealing},  and helps reduce the optical losses due to the waveguide bends \cite{arriola2013low}.
Overall, we were able to achieve transmissivities of up to $(52.6\pm3)\%$ for \SI{22} {mm} long devices, with bending radii of \SI{90} {mm}. We  fabricated several different photonic circuits with the geometry shown in Figure \ref{fig:setup}, and  tuned the power splitting of the directional couplers by modifying their interaction length.
We reconstructed the unitary transformations implemented by the two devices of choice (see Appendix), using methods demonstrated in \cite{brisbanemethod,szameitchara} and subsequent numerical optimization. 
 
\noindent
\subsection*{Holevo Information.}
To analyze the amount of information Bob can gain from a single copy of Alice's state we calculate the Holevo quantity
\begin{equation*}
\chi (m)=-\Tr(\rho \log_2 \rho )+\frac{1}{2^m} \sum_{i=0}^{2^m-1} \Tr(\rho_i \log_2 \rho_i ),
\end{equation*}
where $\rho=\frac{1}{2^m} \sum_{i=1}^{2^m} \rho_i$ and $\rho_i=\sum_{k=0}^{d-1} \bigotimes_{j=1}^{m} R(\frac{k \pi}{d}) \ket{P_{ij}} \times \bra{P_{ij}} R(-\frac{k \pi}{d})$ and $\ket{P_{ij}}=\ket{H}$ when the $j$th bit of $i$ is 0, otherwise $\ket{P_{ij}}=\ket{V}$ \cite{rohde2012quantum}.
In our experiment $m=4$ and $12$, yielding
\begin{equation*}
\chi (4)=1.9694~.
\end{equation*}
Note that for elliptical polarization encodings the Holevo information is halved but the scaling in $m$ remains the same (see Appendix).

\noindent
\subsection*{Bob's random attack.}
 The simplest attack is realized by measuring all photons in the same basis as described in  \cite{rohde2012quantum}. The  probability of inferring the correct state is then given by
\begin{equation}
p=\frac{1}{d} \sum^{d-1}_{j=0} cos^{2m}\left(\frac{j \pi}{d}\right)
\end{equation}
with the number of spatial modes $m$ and the number of possible polarization bases $d$.

\noindent
\subsection*{Measurement errors.}
The main drawback of downconversion-sources is that their emission is probabilistic. This is especially problematic for our experiment, where the probability of simultaneously generating exactly one pair in each crystal, as desired, equals the probability of generating exactly two pairs in one of the crystals. In our setup, we circumvented this problem by making the pairs from the two sources distinguishable by polarization. For input states in which one photon has polarization orthogonal to that of the other three, the input polarization could be set to either $\ket{H,H}$ or $\ket{V,V}$ for source 1, as needed, and $\ket{H,V}$ for source 2: then Alice's final polarization measurement would distinguish the events of interest from those in which one crystal created all four photons. We can also deal with input states with two $\ket{H}$ photons and two $\ket{V}$ photons by having sources 1 and 2 produce $\ket{H,H}$ and $\ket{V,V}$, respectively, and rewiring the input channels to the chip as needed. Double-pair emission for input states $\ket{H,H,H,H}$ and $\ket{V,V,V,V}$ cannot be dealt with this way, but these states are not of interest for a quantum random walk.

Having suppressed errors from double-pair emission, we must now consider triple-pair emission. The noise contributed by these events is on the order of the sources' per-pulse emission probability, which is $0.14\%$. 

To quantify the spectral distinguishability of our photons, we measured Hong-Ou-Mandel interference visibility for all four combinations of signal and idler from source 1 with signal and idler from source 2. After subtracting statistically expected higher-order noise, we measured the visibilities to be
\begin{equation}
V=\frac{C_{max}-C_{min}}{C_{max}}=0.88 \pm 0.05,
\end{equation} 
and $V=0.77\pm 0.05$ without subtracting higher order noise.
For more discussion of experimental errors in quantum random walks, see \cite{tillmann2015generalized}.
We assumed Poissonian error for all single-photon detection rates, so that for $N$ detections, we assume an error of $\epsilon=\sqrt{N}$.

The error in the reconstructed unitary propagates from errors in our intensity measurements, which are in turn used to infer amplitudes and phases. Here we are able to limit the error on the inferred transmission amplitudes and phases to $1\%$ and $\SI{50}{mrad}$, respectively. The discrepancy in error-bar size for the various output possibilities stems from the nature of the unitary: phase errors can lead to large changes in some output probabilities, while having hardly any effect in others.

\noindent
\subsection*{ Unitary of the devices}
Here we present the two unitaries used for the random walk experiments. 

\vspace*{0.5cm}
$
U_1=\left(
\begin{matrix}
0.74&0.38&0.39&0.4\\
0.37&-0.34-0.71i&-0.17+0.31i&-0.18+0.31i\\
0.38&-0.15+0.29i&-0.81+0.06i&0.18+0.25i\\
0.42&-0.17+0.32i&0.2+0.18i&-0.78+0.08i\\
\end{matrix}
\right)$

$
U_2=\left(
\begin{matrix}
0.64&0.44&0.37&0.54\\
0.44&-0.33-0.65i&-0.14+0.26i&-0.15+0.41i\\
0.37&-0.14+0.26i&-0.4+0.51i&-0.15-0.57i\\
0.54&-0.15+0.41i&-0.15-0.57i&-0.41+0.02i\\
\end{matrix}
\right)$

\begin{figure*}[h]
\center
\includegraphics[width=0.8\textwidth]{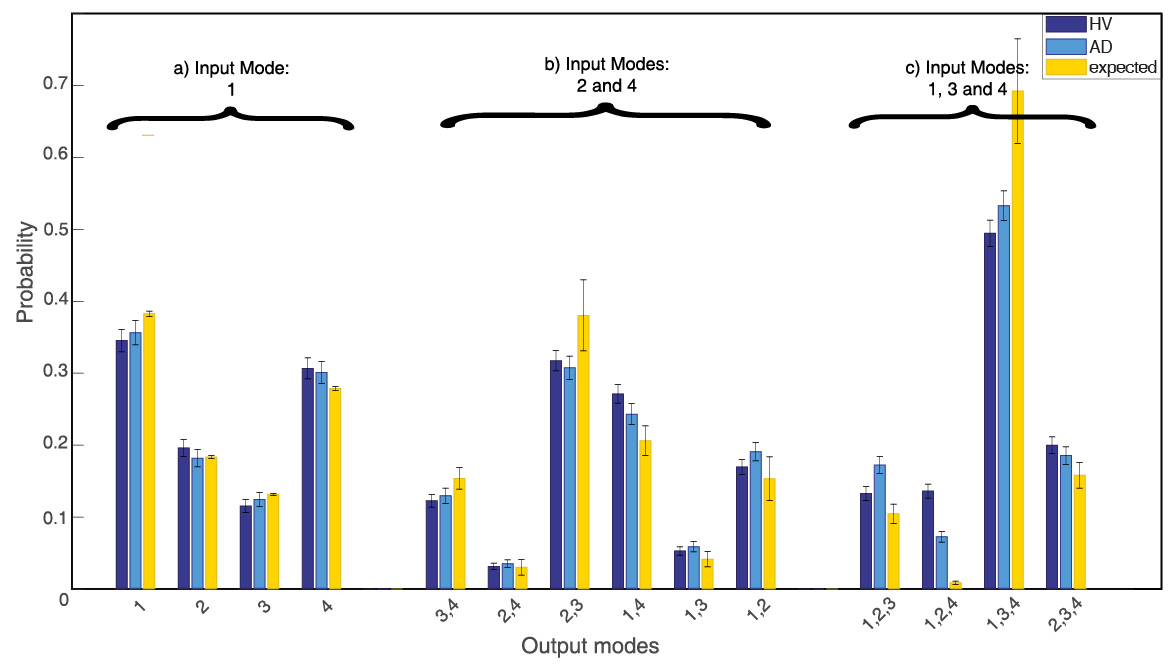}
\caption{\textbf{ Random Walk Results for device 2.}  a) One photon is $H (A)$ polarized and launched into one of the four input modes of the chip while three $V (D)$ polarized photons are send to the other inputs to mask the true input state. The graphs show the probability of finding the $H (A)$ polarized photon in the different output modes. b) Two $H (A)$ polarized photons are launched into the modes 1/2 alongside $V (D)$ polarized photons in the other modes. The probabilities to find the two photons in different output modes are shown. c) Three $H (A)$  photons are launched into the modes 2,3,4 alongside a $V (D)$ photon in the remaining mode. The fidelities (Bhattacharyya distance) between the expected and measured probabilities are (a) $0.99\pm0.02$, (b) $0.99\pm0.02$ and (c) $0.96\pm0.026$.}
\label{fig:results}
\end{figure*}

\noindent
\subsection*{ Holevo information for encryption using random rotations over the Poincare sphere}

Any operation on the Poincare sphere is a SU(2) transformation for which an arbitrary element can be expressed as
\begin{align}
R(\alpha,\beta,\gamma)=R_z(\alpha)R_y(\beta)R_z(\gamma) \ ,	
\end{align}
where $R_z(\alpha)=e^{-i\alpha\sigma_z}$, $R_y(\beta)=e^{-i\beta\sigma_y}$, and
\begin{align}
	\sigma_z=\frac{1}{2}\left(
	\begin{array}{cc}
	1 & 0 \\
	0 & -1
	\end{array}\right ) \ ,
	\sigma_y=\frac{1}{2}\left(
	\begin{array}{cc}
	0 & -i \\
	i & 0
	\end{array}\right ) \ .
\end{align} 
Then explicitly in the basis $\{\ket{H}, \ket{V}\}$,
\begin{align}\label{eq:SU2}
R(\alpha,\beta,\alpha)=	
	\left(
	\begin{array}{cc}
	\cos\left(\frac{\beta}{2}\right)e^{-\frac{i}{2} (\alpha+\gamma)} & -\sin\left(\frac{\beta}{2}\right)e^{-\frac{i}{2} (\alpha-\gamma)} \\
	\sin\left(\frac{\beta}{2}\right)e^{\frac{i}{2} (\alpha-\gamma)} & \cos\left(\frac{\beta}{2}\right)e^{\frac{i}{2} (\alpha+\gamma)}
	\end{array}\right ) \ .
\end{align}
The Haar measure on SU(2) is
\begin{align}
	d\mu(\alpha, \beta,\gamma)=\frac{1}{16\pi^2}\sin(\beta) d\beta d\alpha d\gamma \ .
\end{align}
So, to sample uniformly from SU(2) on the Poincare sphere, we have to pick $\alpha, \gamma\in [0,2\pi)$, and $\xi\in[0,1]$ and compute $\beta=2\sin^{-1}(\sqrt{\xi})$ (see Section 2.3 in essay ``How to generate a random unitary matrix" by Maris Ozols).

\noindent
\subsubsection*{Encryption}
Alice selects secret keys $k_1, k_2$, and $k_3$ at random uniformly, where $k_i\in [0,d_i-1]$. She performs a mode-by-mode random polarization on her input state from the Poincare defined by the (Euler) angles of the transformation:
\begin{align}
	\alpha=\frac{2\pi k_1}{d_1}, \ \gamma=\frac{2\pi k_3}{d_3}, \ \beta=2\sin^{-1}(\sqrt{\xi}) \ {\rm for} \  \xi=\frac{k_2}{d_2 -1}.
\end{align}
If Alice's $m$-mode input state is $\ket{\psi}$, then she implements $R\left(\alpha,\beta,\gamma\right)^{\otimes m}\ket{\psi}$.
As in Rohde et al., each mode of Alice input state is either $\ket{H}$ for logical 0, or $\ket{V}$ for logical 1. Let us denote Alice's encoded bit-string by ${\bf x}$. After encryption, Alice sends her encrypted state to Bob for processing. Since Bob does not know Alice's secret keys, he sees a superposition of all the states corresponding to the different possible secret keys,
\begin{align}
\rho_{\bf x}=\frac{1}{d_1 d_2 d_3}\sum_{k_1=0}^{d_1-1}\sum_{k_2=0}^{d_2-1}\sum_{k_3=0}^{d_3-1}\bigotimes_{j=1}^m R(\alpha,\beta,\gamma)\ket{P_{{\bf x},j}}\bra{P_{{\bf x},j}}R(-\alpha,-\beta,-\gamma) \ ,
\end{align}
where $\ket{P_{{\rm x}, j}}=\ket{H}$ when the $j$th bit of ${\rm x}$ is 0, and $\ket{P_{{\rm x}, j}}=\ket{V}$ when the $j$th bit of ${\rm x}$ is 1.

Let $\rho=\frac{1}{2^m}\sum_{{\rm x}\in \mathbb{Z}_2^m} \rho_{\bf x}$. Then, the Holevo quantity is given by
\begin{align}
\chi  \left( {\rho_{ x}, p_{\bf x}=\frac{1}{2^m}}\right) =S \left( \rho \right)-\sum_{{\bf x}\in \mathbb{Z}_2^m} \frac{1}{2^m}S\left(\rho_{\bf x}\right ),
\end{align}
where $S(\sigma)$ is the von Neumann entropy of density matrix $\sigma$.
As in Rohde et al., the first term is equal to $m$ because $\left\{\bigotimes_{j=1}^m \ket{P_{{\bf x},j}}\ , \ {\bf x}\in\mathbb{Z}_s^m\right\}$ forms a complete set of basis. Owing to the invariance of the von Neumann entropy under unitary transformation, we have
\begin{align}
S(\rho_{\bf x})	=S(\rho_{\bf 0}) \ .
\end{align}
Hence, it suffices to find the eigenvalues of $\rho_{\bf 0}$.
First, we observe that 
\begin{align}
	R(\alpha,\beta,\gamma)\ket{H}^{\otimes m}=\left(\cos\left(\frac{\beta}{2}\right)e^{-\frac{i}{2}(\alpha+\gamma)}\ket{H}-\sin\left(\frac{\beta}{2}\right)e^{-\frac{i}{2}(\alpha-\gamma)}\ket{V}\right ) \ ,
\end{align}
where $e^{-\frac{i}{2}\alpha}$ is a global phase factor. Let $\ket{a_V}_m$ denote a state that is the symmetric sum of all states with $a$ qubits in the $\ket{V}$ state. So $\rho_{\bf 0}$ simplifies to
\begin{align}
	\rho_{\bf 0}=&\frac{1}{d_2 d_3}\sum_{k_2=0}^{d_2-1}\sum_{k_3=0}^{d_3-1}\bigotimes_{j=1}^m \left(\cos\left(\frac{\beta}{2}\right)e^{-\frac{i}{2}\gamma}\ket{H}-\sin\left(\frac{\beta}{2}\right )e^{\frac{i}{2}\gamma}\ket{V}\right ) \left(\cos\left(\frac{\beta}{2}\right)e^{\frac{i}{2}\gamma}\bra{H}-\sin\left(\frac{\beta}{2}\right )e^{-\frac{i}{2}\gamma}\bra{V}\right )\\
	=& \frac{1}{d_2 d_3}\sum_{k_2=0}^{d_2-1}\sum_{k_3=0}^{d_3-1}\sum_{a,b=0}^m \sqrt{\left(\begin{array}{c}m\\ a\end{array}\right ) \left(\begin{array}{c}m\\ b\end{array}\right )} \left(\cos\left(\frac{\beta}{2}\right )e^{-\frac{i}{2}\gamma}\right )^{m-a}  \left(-\sin\left(\frac{\beta}{2}\right )e^{\frac{i}{2}\gamma}\right )^a \\
	&\times \ket{a_V}_m {}_m\bra{b_V} \left(\cos\left(\frac{\beta}{2}\right )e^{\frac{i}{2}\gamma}\right )^{m-b}  \left(-\sin\left(\frac{\beta}{2}\right )e^{-\frac{i}{2}\gamma}\right )^b \ . \label{eq:rho0}
	\end{align}
The sum over $k_3$ is
\begin{align}
&	\frac{1}{d_3}\sum_{k_3=0}^{d_3-1} \exp\left(-\frac{i}{2}\gamma (m-a-a-(m-b)+b)\right )\\
	&=\frac{1}{d_3}\sum_{k_3=0}^{d_3-1} \exp\left(-\frac{i}{2}\gamma (-2a+2b)\right )\\
	&=\frac{1}{d_3}\sum_{k_3=0}^{d_3-1} \exp\left(-i\gamma (-a+b)\right )\\
	&=\frac{1}{d_3}\sum_{k_3=0}^{d_3-1} \exp\left(-\frac{2\pi i k_3}{d_3} (-a+b)\right )\\
	&=\delta_{a,b} \ .
\end{align}
As such, eq.~(\ref{eq:rho0}) simplifies to

\begin{align}
	\rho_{\bf 0}=&\frac{1}{d_2}\sum_{a=0}^m\sum_{k_2=0}^{d_2-1}\left(\begin{array}
{c} m \\ a	
\end{array}
\right) \left(\cos^2\left(\frac{\beta}{2}\right )\right)^{m-a}\left(\sin^2\left(\frac{\beta}{2}\right)\right)^a \ket{a_V}_m {}_m\bra{a_V}\\
=&\frac{1}{d_2}\sum_{a=0}^m\sum_{k_2=0}^{d_2-1}\left(\begin{array}
{c} m \\ a	
\end{array}
\right) \left(\frac{d_2-1-k_2}{d_2-1}\right)^{m-a}\left(\frac{k_2}{d_2-1}\right)^a \ket{a_V}_m {}_m\bra{a_V} .
\end{align}

In the limit of large $d_2$, we can approximate the sum over $k_2$ as an integral, {\it i.e.}
\begin{align}
	\frac{1}{d_2}\sum_{k_2=0}^{d_2-1}\left(1-\frac{k_2}{d_2-1}\right )^{m-a} \left(\frac{k_2}{d_2-1}\right )^{a}\rightarrow \int_0^1 dx (1-x)^{m-a} x^a \ .
\end{align}
Using an identity
\begin{align}
	\int_0^1 t^{x-1}(1-t)^{y-1} dt=\frac{\Gamma(x)\Gamma(y)}{\Gamma(x+y)} \ ,
\end{align}
we have 
\begin{align}
	\int_0^1 dx (1-x)^{m-a} x^a = &\frac{\Gamma(a+1)\Gamma(m-a+1)}{\Gamma(m+2)}\\
	=& \frac{a!(m-a)!}{(m+1)!} \ .
\end{align}
Putting all this together, we have
\begin{align}
	\lim_{d_2\rightarrow\infty}\rho_{\bf x}=&\sum_{a=0}^m \left(\begin{array}
{c} m\\ a	
\end{array}
\right ) \frac{a!(m-a)!}{(m+1)!}\ket{a_V}_m {}_m \bra{a_V} \\
=& \frac{1}{m+1}\sum_{a=0}^m \ket{a_V}_m {}_m \bra{a_V} \ .
\end{align}
When $d_2\rightarrow\infty$, $\rho_{\bf 0}$ is completely mixed over the symmetrized basis states $\ket{a_V}_m$ and its entropy, which represents the number of bits hidden from Bob, is $\log_2(m+1)$. For linear polarization  the entropy of the encrypted all-zero bit string is $\frac{1}{2}\log_2(\pi em /2)$. Hence, the randomization over the whole Poincare sphere decreases the Holevo information by a constant factor, but the scaling in $m$ remains $\mathcal{O}(\log_2(m))$.

\end{document}